# Investigation of real-fluid effects on $NH_3$ oxidation and blending characteristics at supercritical conditions via high-order Virial equation of state coupled with *ab initio* intermolecular potentials

Mingrui Wang[a], Ruoyue Tang[a], Xinrui Ren[a], Hongqing Wu[a], Yuxin Dong[a], Ting Zhang[a], Song Cheng[a,b*]

[a] *Department of Mechanical Engineering, The Hong Kong Polytechnic University, Kowloon, Hong Kong SAR, China*
[b] *Research Institute for Smart Energy, The Hong Kong Polytechnic University, Kowloon, Hong Kong SAR, China*

*Corresponding author.
Email: songcheng@polyu.edu.hk
Song Cheng
Phone: +852 2766 6668

# Abstract

Significant efforts have been committed to understanding the fundamental combustion chemistry of ammonia at high-pressure and low-temperature conditions with or without blending with other fuels, as these are promising to improve ammonia combustion performance and reduce NOx emission. A commonly used fundamental reactor is the jet-stirred reactor (JSR). However, modeling of high-pressure JSR experiments have been conducted assuming complete ideal gas behaviors, which might lead to misinterpreted or completely wrong results. Therefore, this study proposes, for the first time, a novel framework coupling high-order Virial equation of state, *ab initio* multi-body intermolecular potential, and real-fluid governing equations. The framework is further applied to investigate $NH_3$ oxidation under supercritical conditions in jet-stirred reactors, where the real-fluid effects on $NH_3$ oxidation characteristics are quantified and compared, via simulated species profiles and relative changes in simulated mole fractions, at various temperatures, pressures, dilution ratios, equivalence ratios, and with or without blending with $H_2$ and $CH_4$. Strong promoting effects on $NH_3$ oxidation from real-fluid effects are revealed, with significant shifts in simulated species profiles observed for both fuel, intermediates and product species. Sensitivity analyses are also conducted based on the new framework, with diverse influences of real-fluid effects on the contributions of the most sensitive pathways highlighted. It is found that, without considering real-fluid behaviors, the error introduced in simulated species mole fractions can reach ±85% at the conditions investigated in this study. Propagation of such levels of error to chemical kinetic mechanisms can disqualify them for any meaningful modeling work. These errors can now be excluded using the framework developed in this study.

Keywords: high-order Virial equation of state; *ab initio* intermolecular potential; ammonia; supercritical oxidation; jet-stirred reactor

# Novelty and Significance Statement

This study is the first time that high-order Virial equation of state, *ab initio* multi-body potentials and real-fluid conservation laws are applied to supercritical $NH_3$ oxidation modelling in jet-stirred reactors. This is enabled through a newly developed framework that adequately represents fluid nonidealities. Through numerical investigation on $NH_3$ oxidation, we reveal and quantify the real fluid effects on supercritical oxidation processes, and compare these effects over a wide range of temperatures, pressures, dilution ratios, equivalence ratios, and with or without fuel blending with $H_2$ and $CH_4$. Furthermore, the significant error introduced to chemical kinetic models by assuming ideal gas behaviors during validation against high-pressure JSR experiments is characterized, which can now be eliminated by the framework proposed in this study.

# 1. Introduction

The increasing concerns over greenhouse gas emissions have intensified the demand for the application of carbon-neutral energy. Searching for alternative fuels with low or zero $CO_2$ emission and high combustion efficiency is one of the most significant research hotspots. In addition to carbon-based fuels such as $CH_4$, $CH_3OH$, and $C_2H_4$, ammonia ($NH_3$) is recognized as a promising alternative fuel owing to its carbon-free characteristics. Compared with hydrogen, ammonia can be stored, transported, and liquified at a higher temperature, and has a higher energy density of 20.5 $MJ/m^3$ (for $H_2$, the value is 8.5 $MJ/m^3$) [1]. Furthermore, with green synthesis technology, ammonia can be produced by adopting electricity from renewable energy, e.g., wind power, solar power, and hydropower [2]. Therefore, $NH_3$ is expected to have a wide application in the transportation and power generation industries [3, 4].

Despite the advantages of ammonia, there are many challenges confronting the use of ammonia as a fuel. On the one hand, compared with traditional hydrocarbon fuels, ammonia has lower chemical reactivity, narrow flammability range, and slow flame propagation [5], thereby making it difficult to ignite and easily leading to incomplete combustion. To address this issue, hydrogen and methane are usually served as active fuel additives to blend with $NH_3$ to promote oxidation and ignition reactivity for $NH_3$ [6]. On the other hand, nitrogen oxides (mainly NO, $N_2O$, $NO_2$), which are harmful to human health, environment, and climate, are abundantly produced during oxidation and combustion of $NH_3$ [7]. Except for adopting fuel blending to lower the formation of nitrogen oxides [8], low temperature [9] and high pressure [10] combustion have been used to suppress NOx formation.

With such awareness, to improve oxidation characteristics, and fulfill the requirement of high energy density in the transportation area, low-temperature high-pressure ammonia oxidation with proper additives is a promising topic [11]. As on-road heavy-duty engines, gas turbines, and rocket engines can operate up to 300 atm [12], higher than the critical pressure of $NH_3$ (113 atm) and related mixtures, supercritical combustion is commonly observed, where intermolecular interaction becomes unignorable. Different from the ideal gas, supercritical fluids exhibit unique features, e.g., high density, high diffusivity [13], and high compressibility factors [14]. The shift in thermochemical properties has a profound influence on combustion characteristics, e.g., ignition temperature, ignition delay time, and flame propagation speed [15]. Thus, it's necessary to get an accurate and quantitative description of the real-fluid effect on supercritical oxidation of $NH_3$ at low temperatures, which is, however, absent from previous studies.

To develop an accurate model for $NH_3$ oxidation, with the provision of uniform mixing, precise control of temperature and pressure, and steady-state operation, jet-stirred reactors (JSRs) are widely adopted for research. To date, limited by the facility structure, ammonia oxidation in a JSR is mainly investigated at around 1 atm [16-19]. Recently, an experimental study on an ammonia/n-heptane mixture was conducted up to 100 bar in a JSR, with an updated chemical kinetic model developed [20]. However, there are still some limitations for these experiment-based investigations: (1) existing results are not enough to systematically investigate the oxidation characteristics of $NH_3$ at various supercritical conditions; (2) to consider the real-fluid effect, the fuel chemistry model is traditionally developed against indirect experimental measurements (e.g., species profiles and ignition delay times) while ideal-gas thermochemistry is still used, leading to uncertainties caused by the real-fluid effect. Previous studies [21, 22] have demonstrated that the prediction results based on these chemistry models show high sensitivity to these uncertainties. If these chemistry models are utilized at a pressure higher than the experimental pressure, the uncertainties are significant and influence the accuracy. However, these uncertainties are conventionally ignored and seldom investigated in previous studies at a wide condition range, limiting the utilization and improvement of these chemistry models.

For these reasons, numerical investigations are preferred and are conducted in this research. Except for modifying chemical kinetics, correcting thermochemical properties is another way to incorporate the nonideal behavior into a combustion model [23]. Conventionally, the real-fluid effect is considered by adopting various equations of state (EoS), typically the cubic EoS, such as Redlich- Kwong (RK) EoS [24], Peng-Robinson EoS [25, 26], and Soave-Redlich-Kwong EoS [27]. These previous numerical investigations still have the following insufficiencies: (1) despite the simplicity and convenience, cubic EoS is empirical and thus cannot reveal the essence of the real-fluid effect from the fundamental viewpoints of intermolecular interactions; (2) the supercritical oxidation of $NH_3$ in a JSR hasn't been investigated systematically, and thus most previous studies are unable to guide the selection of operation conditions for $NH_3$ oxidation. Recently, high-order Virial EoS, the parameters of which are directly derived from molecular potentials, have been successfully applied in supercritical oxidation and autoignition simulations in JSRs and rapid compression machines/shock tubes, respectively, by the authors' group [28, 29]. In these studies, the high-order Virial EoS coupled with the *ab initio* two- to multi-body intermolecular potentials has been demonstrated to outperform the traditional cubic EoS and ideal EoS considerably, predicting nearly identical results with experimental measurements for the thermodynamic properties of supercritical mixtures.

Therefore, this study aims to develop a framework that can adequately replicate the real-fluid behaviors during supercritical oxidation in JSRs, and based on which, the real-fluid effects on the oxidation of $NH_3$ are further characterize at wide ranges of temperature, pressure, equivalence ratio, dilution ratio, and with or without fuel blending with $H_2$ or $CH_4$. The error and misinterpretation introduced by modeling high-pressure JSR experiments with ideal gas assumption are also quantified and discussed. The remaining manuscript is organized as follows. Section 2 provides a description of the methodologies developed in this study. Section 3 discusses the results of this work. This is followed by a summary of the paper in Section 4.

## 2. Methodologies

### 2.1 High-order mixture Virial EoS

The $N^{\text{th}}$-order Virial equation of state is stated as:

$$Z = \frac{P\bar{v}}{RT} = \frac{\bar{v}}{\bar{v}^{IG}} = 1 + \frac{B_2}{\bar{v}} + \cdots + \frac{B_N}{\bar{v}^{N-1}} \tag{1}$$

where $Z$, $P$, $T$, $\bar{v}$, and $R$ denote the compressibility factor, pressure, temperature, molar volume, and universal gas constant, respectively. The second, third, …, and $N^{\text{th}}$-order Virial coefficients are represented by $B_2$, $B_3$, …, and $B_N$, related to intermolecular interactions between two molecules, three molecules, …, and $N$ molecules, respectively.

When Virial EoS is used for mixtures or pure substances, the related $B_N$ is accordingly named mixture Virial coefficients or pure-substance Virial coefficients, and the former one can be calculated by the mixing rule [30]. For example, for an R-S binary mixture, the $N^{th}$-order mixture Virial coefficient is calculated by:

$$B_N(T, X_R, X_S) = \sum_{r+s=N;\, r,s \geq 0} \frac{N!}{r!\, s!} X_R^r X_S^s B_{rs} \qquad (2)$$

where $B_{rs}$ represents the Virial coefficient for $r$ molecules of component $R$ and $s$ molecules of component $S$, while $X_R$ and $X_S$ are mole fractions of each substance. If $r = 0$ or $s = 0$, $B_{rs}$ represents the $N^{th}$-order pure-substance Virial coefficients of the substance $R$ or $S$, respectively. If $r \neq 0$ and $s \neq 0$, $B_{rs}$ denotes the $N^{th}$-order cross Virial coefficient, corresponding with the intermolecular interaction between components $R$ and $s$.

To date, based on *ab initio* potentials, pure-substance Virial coefficients and cross Virial coefficients have been derived for several species (listed in Table S1 in the Supplementary Material), where these species usually account for high mole fractions in a combustion system (e.g., oxidizers, fuels, and final combustion products). However, there are usually hundreds of other substances that can be formed during oxidation, particularly short-lived radicals and intermediates. Considering the computation cost, it is expensive to calculate the *ab initio* 2$^{nd}$- to $N^{th}$-order Virial coefficients from two-body to multi-body molecular potentials for all the species involved in a combustion process. For instance, to compute the 2$^{nd}$-order mixture Virial EoS for the mechanism GRI-Mech 3.0 [31], we need to determine 53 pure-substance 2$^{nd}$-order Virial coefficients and 1378 2$^{nd}$-order cross Virial coefficients, with each coefficient requiring thousands of *ab initio* calculations. To address this issue, recently, based on *ab initio* molecular potentials, our group proposed a framework to derive a set of high-order mixture Virial EoS for an arbitrary system [29]. Note that *ab initio* intermolecular potentials are used for all the major species in this study, while the framework is used to derive the Virial coefficients for radicals and other minor species that impose relatively less influences on overall mixture behaviors. The framework is mainly composed of three steps (Fig. 1) and is briefly illustrated here.

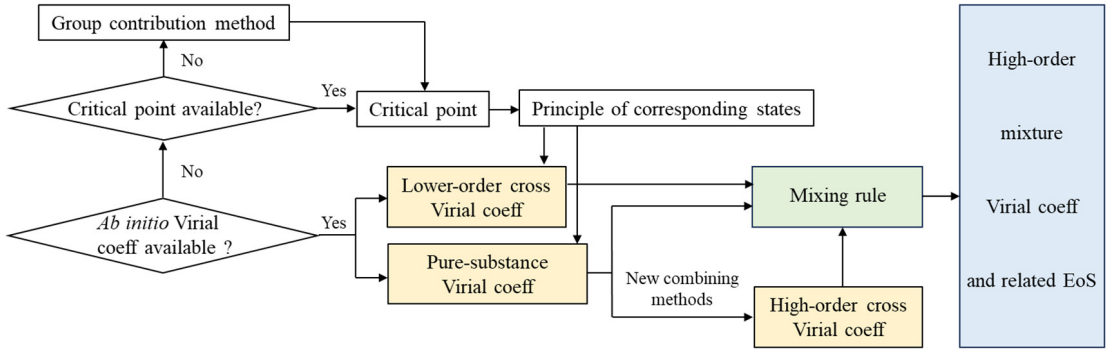

**Figure 1. The framework for deriving high-order mixture Virial EoS**

### 2.1.1 Determining pure-substance Virial coefficients

Except for those derived before (Table S1), the 2$^{nd}$- and 3$^{rd}$-order Virial coefficients of pure substances, including intermediates and radicals, are estimated by Tsonopoulos [32] (Eq. 3) and Orbey-Vera [33] (Eq. 4) correlations, respectively, which are based on the principle of corresponding state, expressing the Virial coefficient as a function of critical temperature and pressure.

$$\frac{B_2 P_{c,i}}{RT_{c,i}} = f_0(T_r) + \omega_i f_1(T_r) + a(\mu_i) f_2(T_r) \qquad T_r = \frac{T}{T_{c,i}} \qquad (3)$$

$$\frac{B_3 P_{c,i}^2}{(RT_{c,i})^2} = f_3(T_r) + \omega_i f_4(T_r) \qquad (4)$$

where, $P_{c,i}$, $T_{c,i}$ are respectively the critical pressure and temperature of substance $i$, $\omega_i$ is the acentric factor, $a$ is a function of the dipole moment $\mu_i$, and $f_0 - f_4$ are functions of the dimensional temperature $T_r$ [32, 33].

If necessary, the Joback group contribution method [34] (JGC) is employed to estimate critical temperatures and pressures. Notably, the JGC method is only suitable for stable species. Consequently, radicals should be regarded as analogous stable species when estimating critical properties, similar to the treatment in previous studies [35, 36].

### 2.1.2 Estimating low-order cross Virial coefficients

Except for those in Table S1, the 2$^{nd}$-order cross Virial coefficients are determined still using the Tsonopoulos correlations [32], while the critical point properties, acentric factors, and the dipole moments in Eq. 3 should be replaced with the corresponding binary mixture properties, which are calculated based on non-ideal mixing rules [28], following:

$$T_{c,ij} = (1 - k_{ij})\sqrt{T_{c,i} T_{c,j}} \qquad (5)$$

$$P_{c,ij} = \frac{4RT_{c,ij}(Z_i + Z_j)}{\left(\bar{v}_{c,i}^{1/3} + \bar{v}_{c,j}^{1/3}\right)^3} \qquad (6)$$

$$\omega_{ij} = 0.5(\omega_i + \omega_j) \qquad (7)$$
$$\mu_{ij} = 0.5(\mu_i + \mu_j) \qquad (8)$$

where, $k_{ij}$ is a binary interaction parameter given by:

$$k_{ij} = 1 - \frac{8(\bar{v}_{c,i} \bar{v}_{c,j})^{0.5}}{\left(\bar{v}_{c,i}^{\frac{1}{3}} + \bar{v}_{c,j}^{\frac{1}{3}}\right)^2} \qquad (9)$$

The 3$^{rd}$-order cross Virial coefficients are computed using Chueh-Prausnitz [37] correlations. Taking an N$_2$-O$_2$-Ar mixture as an example, the 3$^{rd}$-order cross Virial coefficient $B_{111}$ is written as:

$$B_{111} = \sqrt[3]{C_{110} C_{101} C_{011}} \qquad (10)$$

where, $C$ is named the 'pseudo third Virial coefficient' of a binary system calculated by Eq. 4. For instance, $C_{110}$ represents one molecule of N$_2$ and one molecule of O$_2$, calculated by using N$_2$-O$_2$ mixture properties that have been determined *ab initio*.

### 2.1.3 Estimating high-order cross Virial coefficients

High-order ones are directly predicted from *ab initio* high-order pure-substance Virial coefficients using the "new combining method" [29] proposed recently. For instance, the expression of the $(r+s)^{th}$-order cross Virial coefficient for a binary *R-S* system is:

$$B_{rs} = \frac{rB_{r+s,0} + sB_{0,r+s}}{r+s} \qquad r,s \geq 0 \qquad (11)$$

where, $B_{r+s,0}$ and $B_{0,r+s}$ are the $(r+s)^{th}$-order pure-substance Virial coefficients of substances R and S, respectively. This new combining method can be extended to any multi-component system, which has been validated previously [29].

### 2.1.4 Calculating mixture Virial coefficients

By the mixing rule (Eq. 2), mixture Virial coefficients are calculated by cross Virial coefficients and pure-substance Virial coefficients, subsequently generating high-order mixture Virial EoS. Compared with ideal and RK EoS, the derived high-order mixture Virial EoS has been proven to have a higher accuracy in predicting thermochemical properties of various gases over wide range of conditions. For detailed information, please refer to Chapter 3 of the Supplementary Material.

## 2.2 JSR model based on high-order Virial EoS

### 2.2.1 Governing equations

A recent study [28] from the authors' group has derived and validated the governing equations of real-fluid oxidation in a JSR based on a high-order mixture Virial EoS, where its superiority over traditional ideal EoS and RK EoS is demonstrated. The derivation is briefly introduced here. For an *I*-component system, several assumptions are made:

(1) The steady-state flow has uniform properties in the direction perpendicular to the flow.
(2) The steady-state pressure and temperature are directly given.

The mass conservation of the *i*th species in the reactor can be written as:

$$\dot{m}_{cv} = V M_i w_i + \dot{m}_{in} Y_{i,in} - \dot{m}_{out} Y_{i,out} \qquad i = 1,2\ldots, I\ species \qquad (12)$$

where $M_i$, $w_i$ and $Y_{i,in}$ are the molecular weight, net production rate, and inlet mass fraction of the $i$ species, respectively. $V$ is the volume of the reactor. $\dot{m}_{cv}$ is the mass change rate in the reactor, and $\dot{m}_{in}$ and $\dot{m}_{out}$ are the mass flow rates at the inlet and outlet, respectively, of the reactor. For steady-state flow,

$$\dot{m}_{cv} = 0, \dot{m}_{in} = \dot{m}_{out} = \dot{m} \qquad (13)$$

In addition, the mole fraction $X_i$ and mass fraction $Y_i$ obey:

$$Y_i = \frac{X_i M_i}{\overline{M}} \qquad (14)$$

where $\overline{M}$ is the average molar mass in the reactor. Combining Eqs. 12-14, the mass conservation for each species becomes:

$$f_{mas,i}(\bar{v}, X_1 \ldots X_I) = \frac{V w_i}{\dot{m}} + \frac{X_{i,in}}{\overline{M}_{in}} - \frac{X_i}{\overline{M}} = 0 \qquad i = 1,2\ldots, I - 1\ species \qquad (15)$$

According to the Gibbs Phase Rule [38], the steady state of this system is determined by temperature, molar volume, and the mole fraction of each substance. As the temperature is constant and has been given, there are $I + 1$ unknown variables ($\bar{v}, X_1 \ldots X_I$) that need to be solved, while there are only $I - 1$ mass conservation equations given by Eq. 15 and another one from the condition that the sum of all mole fractions must be 1 (Eq. 17). Therefore, the $N^{th}$-order mixture Virial EoS is used as another governing equation (Eq. 16).

$$f_{eos}(\bar{v}, X_1 \ldots X_I) = -\frac{P\bar{v}}{RT} + 1 + \sum_k \frac{B_k}{\bar{v}^{(k-1)}} = 0 \qquad k = 2,3,\ldots N \qquad (16)$$

$$f_{sum}(\bar{v}, X_1 \ldots X_I) = \sum_i X_i - 1 = 0 \qquad (17)$$

Eqs. 15-17 compromise the final set of governing equations, which are $I + 1$ algebraic equations, representing a classical two-point boundary value problem.

## 2.2.2 Root-finding algorithm

Similar to the numerical method in simulating ideal gas oxidation in a JSR [39], the Newton-Raphson method [40] is adopted in this study to find the steady-state solutions for governing equations Eqs. 15-17 For faster and better convergency, a $(I + 1) \times (I + 1)$ Jacobian matrix is indispensable, following:

$$\text{Jacobian matrix} = \begin{bmatrix} \frac{\partial f_{eos}}{\partial \bar{v}} & \frac{\partial f_{eos}}{\partial X_1} & \cdots & \frac{\partial f_{eos}}{\partial X_l} & \cdots & \frac{\partial f_{eos}}{\partial X_I} \\ \frac{\partial f_{mas,1}}{\partial \bar{v}} & \frac{\partial f_{mas,1}}{\partial X_1} & \cdots & \frac{\partial f_{mas,1}}{\partial X_l} & \cdots & \frac{\partial f_{mas,1}}{\partial X_I} \\ \cdots & \cdots & \cdots & \cdots & \cdots & \cdots \\ \frac{\partial f_{mas,i}}{\partial \bar{v}} & \frac{\partial f_{mas,i}}{\partial X_1} & \cdots & \frac{\partial f_{mas,i}}{\partial X_l} & \cdots & \frac{\partial f_{mas,i}}{\partial X_I} \\ \cdots & \cdots & \cdots & \cdots & \cdots & \cdots \\ \frac{\partial f_{mas,I-1}}{\partial \bar{v}} & \frac{\partial f_{mas,I-1}}{\partial X_1} & \cdots & \frac{\partial f_{mas,I-1}}{\partial X_l} & \cdots & \frac{\partial f_{mas,I-1}}{\partial X_I} \\ 0 & 1 & \cdots & 1 & \cdots & 1 \end{bmatrix} \quad (18)$$

where $l = 1,2 \dots, I\ species$. The partial differentials in the Jacobian matrix are also derived, given by:

$$\frac{\partial f_{eos}}{\partial \bar{v}} = -\frac{P}{RT} - \sum_k \frac{(k-1)B_k}{\bar{v}^k} \quad (19)$$

$$\frac{\partial f_{eos}}{\partial X_l} = \sum_k \left( \frac{1}{\bar{v}^{(k-1)}} \frac{\partial B_k}{\partial X_l} \right) \quad (20)$$

$$\frac{\partial f_{mas}}{\partial \bar{v}} = -\frac{V}{\dot{m}\bar{v}^2} \frac{\partial w_i}{\partial \rho_n} \quad (21)$$

$$\frac{\partial f_{mas,i}}{\partial X_l} = \begin{cases} \frac{V}{\dot{m}} \frac{\partial w_i}{\partial X_i} - \frac{\overline{M} - X_i M_i}{\overline{M}^2} & l = i \\ \frac{V}{\dot{m}} \frac{\partial w_i}{\partial X_l} + \frac{X_i M_l}{\overline{M}^2} & l \neq i \end{cases} \quad (22)$$

where $\rho_n$ represents the molar density. The value of $w_i$ and its derivatives $\partial w_i / \partial \rho_n$ and $\partial w_i / \partial X_l$ are calculated by the CANTERA platform [41].

Previous investigations have demonstrated that inlet conditions ($\bar{v}^0, X_1^0 \dots X_I^0$) might not be sufficient as the Newton-Raphson solver and the governing equations are sensitive to initial values [39, 42]. Therefore, inlet conditions are first inputted into CANTERA to determine equilibrium mixture properties ($\bar{v}^E, X_1^E \dots X_I^E$) based on ideal-fluid thermochemistry. Subsequently, the equilibrium properties are used as the initial values for the Newton-Raphson solver. For better understanding, the supercritical JSR modeling framework is illustrated in Fig. 2, which is composed of a "Virial Module", a "CANTERA Module", and a "Solver Module". The supercritical JSR modeling framework is included as a module of the UHPC-RF-Master software package, hosted at the UHPC laboratory website (https://uhpc-lab.org/downloads/).

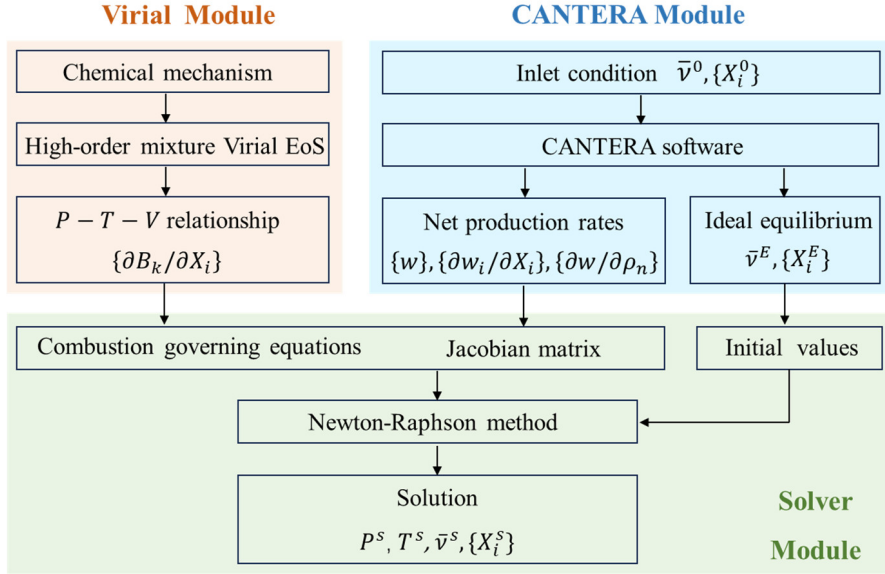

Figure 2. Supercritical oxidation modeling framework based on multi-body intermolecular potential and high-order mixture Virial EoS for jet-stirred reactors.

### 2.2.3 Chemical kinetic mechanism

As mentioned in Fig. 2, a suitable chemical kinetic mechanism is required as the starting point. In this research, the Glarborg-2018 mechanism [43] is utilized for oxidation simulations for $H_2$, $CH_4$, $NH_3$, and related mixtures.

### 2.2.4 Inlet conditions

To systematically explore the real-fluid effects and the contributing factors, simulations are conducted for five different fuel mixtures at a wide range of inlet conditions, as shown in Table 1. The blending ratio (BR) is defined as the molar percentage of $H_2$ or $CH_4$ in the fuel $NH_3$. $CO_2$ is selected as the diluent throughout the investigations, while oxygen is selected as an oxidizer. Cases 1-3 are utilized to compare oxidation characteristics between different fuel mixtures, which are discussed in Sections 3.1 and 3.2. Case 3 is also adopted to explore the influence of temperature, pressure, dilution ratios (D), and equivalence ratios ($\phi$) on oxidation characteristics, which are in Sections 3.2-3.4. Furthermore, Cases 4 and 5 (BR=50%) serve as a comparison with Cases 1-3 to investigate the fuel blending characteristics with $H_2$ or $CH_4$ under the influences of real-fluid effects at various temperatures and pressures (Section 3.5).

Table 1 Inlet conditions for oxidation simulation in supercritical JSR.

| Case | Pressure (bar) | Temperature (K) | Fuel | BR (%) | D (%) | $\phi$ |
|---|---|---|---|---|---|---|
| 1 | 100-1000 | 500-1100 | $H_2$ | 100 | 90 | 0.5 |
| 2 | 100-1000 | 500-1100 | $CH_4$ | 100 | 90 | 0.5 |
| 3 | 100-1000 | 500-1100 | $NH_3$ | 0 | 30, 60, 90 | 0.5, 1, 2 |
| 4 | 100-1000 | 500-1100 | $NH_3/H_2$ | 50 | 90 | 0.5 |
| 5 | 100-1000 | 500-1100 | $NH_3/CH_4$ | 50 | 90 | 0.5 |

# 3. Results and discussion

## 3.1 Compressibility factors

From governing equations Eqs. 15-17, the real-fluid P-T-V relationship plays a significant role in a reacting system while the deviation of it from the ideal-fluid one is quantified by the compressibility factor. Figure 3 compares the compressibility factors of $CO_2$ and three mixtures at T=500-1100 K and P=1-1000 bar, calculated using the high-order mixture Virial EoS developed in Section 2.1.

The diluent $CO_2$ comprises the major part (90%) of all mixtures, thereby dominating the mixture thermochemical properties, resulting in similar compressibility factor distributions for three mixtures and pure $CO_2$, as seen in Fig. 3. The compressibility factor can also be calculated by the ratio of the real-fluid molar volume to the ideal-fluid one (Eq. 1). The compressibility factors of an ideal gas is 1, whereas for real fluids, the values exhibit a strong dependency on temperature and pressure. The compressibility factors for all systems investigated are lower than 1 at a range around P<300 bar and T<650 K. At the other conditions, the compressibility factors are higher than 1 and increase with temperature decreasing or pressure declining. These variations in properties are expected to cause different oxidation characteristics at low and high temperatures.

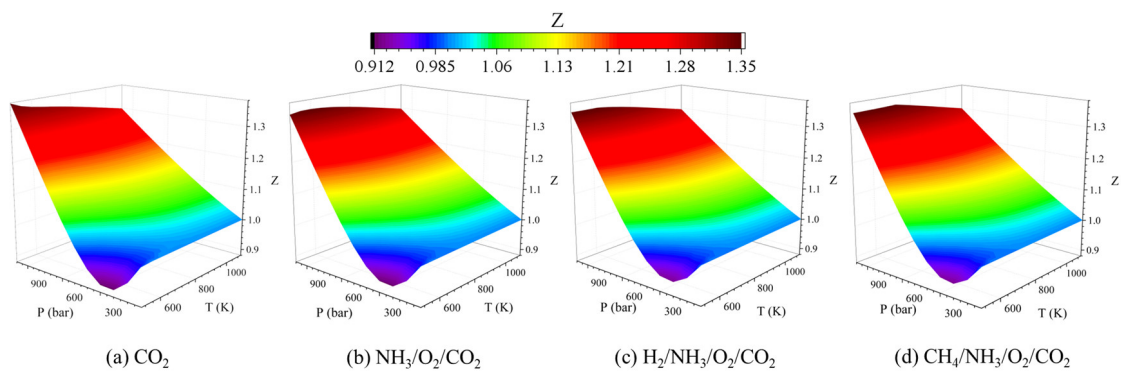

(a) $CO_2$    (b) $NH_3/O_2/CO_2$    (c) $H_2/NH_3/O_2/CO_2$    (d) $CH_4/NH_3/O_2/CO_2$

**Figure 3. Compressibility factors of $CO_2$ and three mixtures (Cases 3-5) at a wider range of temperatures and pressures calculated using high-order mixture Virial EoS.**

## 3.2 Influences of temperatures and pressures

To gain an intuitive insight into the real-fluid effect on oxidation characteristics, Figure 4 compares simulated species profiles based on ideal EoS and high-order Virial EoS for three different mixtures (i.e., Cases 1 – 3 in Table 1) at P=500 bar, dilution ratio of 90% and equivalence ratio of 0.5. For all mixtures investigated, both ideal and Virial EoS show similar trends in mole fraction change with temperature, while there is a distinct gap between them at moderate and high temperatures. With real-fluid effects incorporated, the consumption of fuel and other reactants is promoted, leading to the advanced formation of intermediate species and products. Compared with $CH_4$ (Fig. 4b) and $H_2$ (Fig. 4c), $NH_3$ (Fig. 4a) is more strongly influenced by the real-fluid effect.

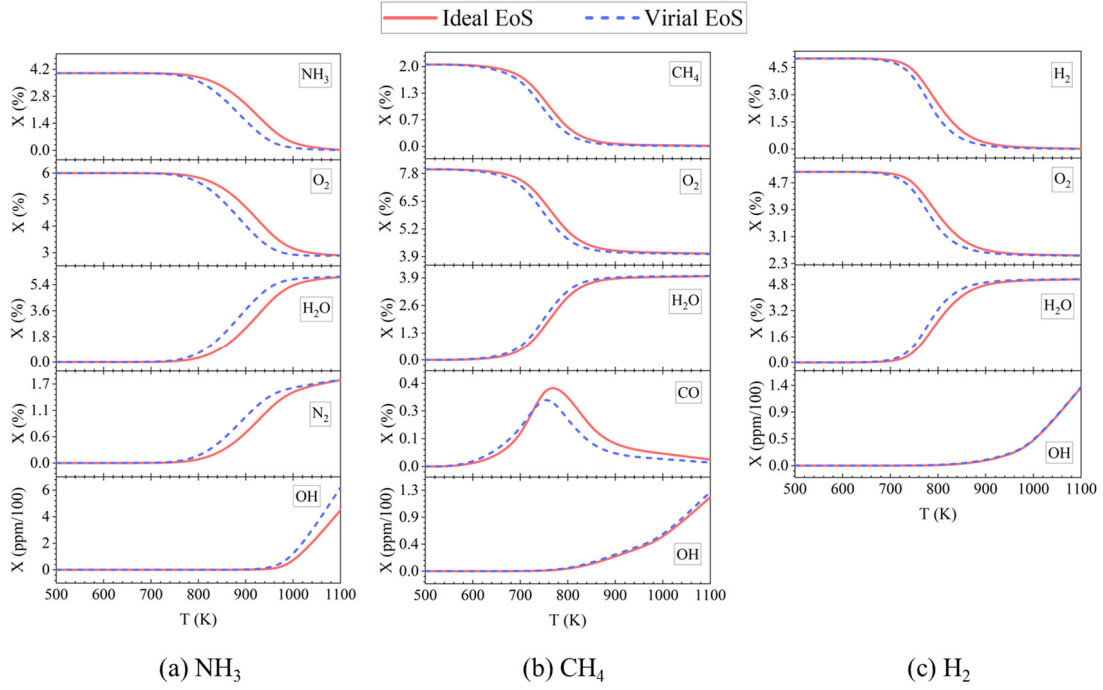

(a) NH$_3$            (b) CH$_4$            (c) H$_2$

**Figure 4. Simulated species profiles during the oxidation of three mixtures (Cases 1-3) based on high-order Virial EoS and ideal EoS at P=500 bar, T=500-1100 K, D=90%, and Φ=0.5.**

The large and non-monotonic changes in supercritical oxidation species profiles in Fig. 4 necessitate investigations of real-fluid effects over wider temperatures and pressures. The real-fluid effects are quantified by $(X_{IG} - X_{VR})/X_{IG}$, which is the normalized change in simulated mole fractions based on Virial EoS from those based on the ideal EoS. A positive (negative) value denotes suppressed (promoted) production or promoted (suppressed) consumption of a corresponding component. A higher absolute value of the normalized change represents a stronger real-fluid effect. The normalized changes for six substances during NH$_3$ oxidation (i.e., Cases 3 in Table 1) are illustrated by 2-D contours in Fig. 5. As the small values of $X_{IG}$ tend to give unphysical values for normalized changes, the contours are truncated when mole fractions are less than 1% of the maximum value of the related species.

Figure 5 clearly demonstrates the significant impact of the real-fluid effect, with a strong dependence on temperature (non-monotonous) and pressure (relatively monotonous). Overall, high pressure augments the real-fluid effects, agreeing well with the findings in [28]. The real-fluid effects are found to improve mixture reactivity and advance the onset of oxidation, resulting in higher consumption of reactants NH$_3$ and O$_2$. These effects are ignorable at T<750 K, where oxidation barely starts (referring to the NH$_3$ profile in Fig. 4(a)) but reaches the highest at around 1000 K, leading to around 75% higher consumption of NH$_3$. For final products, e.g., N$_2$ and H$_2$O, the real-fluid effects promote their production, with stronger effects observed at relatively lower temperatures. The real-fluid effects peak at around T=800 K with approximately 80% higher production. The difference in the mole fractions of these products becomes ignorable at around T=1100 K as the oxidation has nearly completed at this stage. Reversed real-fluid effects are observed for the formation of NO at around 700 K<T<850 K, where the real-fluid effect promotes NO production at lower temperatures while inhibiting NO production at higher temperatures, with the normalized change in NO mole fraction increasing from approximately -18% to 7% with temperature rising from 700 to 950 K. This characteristic for NO, which is referred to as the "reversion phenomenon" herein, is also observed when pressure passes the range around 300 bar<P<500 bar. For NO$_2$, the real-fluid effect promotes its production at all conditions, reaching the highest at around T=700 K, where approximately 80% more NO$_2$ is produced.

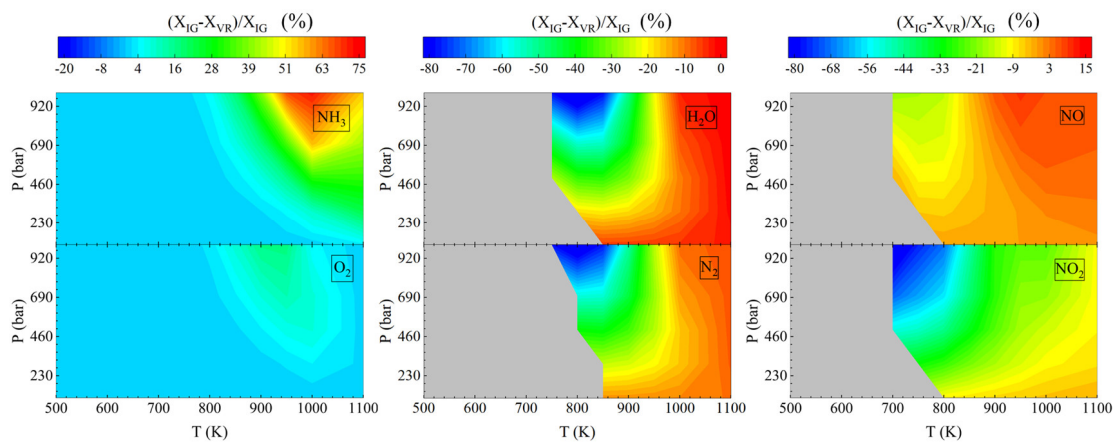

**Figure 5.** Deviation of simulated mole fractions during NH$_3$ oxidation (Case 3) calculated based on high-order Virial EoS from those calculated based on ideal EoS in a supercritical JSR at P=100-1000 bar, T=500-1100 K, D=90% and Φ=0.5.

## 3.3 Influences of dilution ratios

Although the diluent CO$_2$ does not participate in the reaction, the concentration of CO$_2$ considerably affects the thermochemical properties of the reacting mixture (e.g., heat capacity). As such, the influence of diluent ratios on the real-fluid effect on NH$_3$ oxidation is further investigated. Figure 6 records normalized species profiles during NH$_3$/O$_2$/CO$_2$ oxidation (i.e., Cases 3 in Table 1) at two different dilution ratios (i.e., 30% and 90%). The ratio $X/X_{NH3\_0}$ is the ratio of the mole fraction of a component to the initial mole fraction of ammonia, denoting how many products are generated or how many reactants are consumed per mole of NH$_3$ fed to the JSR. It is clear from Fig. 6 that a lower dilution ratio accelerates the oxidation process. This leads to earlier consumption of reactants (NH$_3$ and O$_2$) and earlier production of products (e.g., N$_2$ and H$_2$O) at lower temperatures. For NO and NO$_2$, a lower dilution ratio leads to slightly higher production at relatively low temperatures, e.g., around 700 K<T<900 K; however, at higher temperatures, their production is inhibited at the lower dilution ratio.

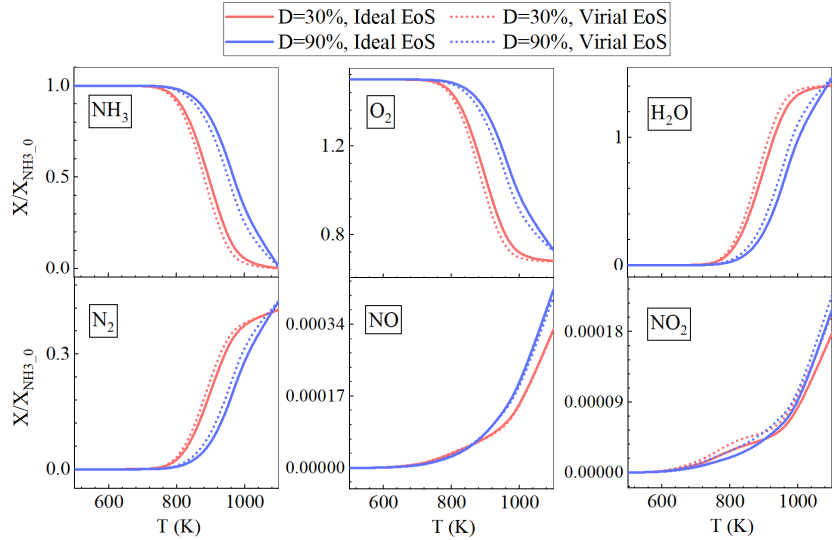

**Figure 6.** Normalized species profiles during NH$_3$/O$_2$/CO$_2$ oxidation (Case 3) based on high-order Virial EoS and ideal EoS in a supercritical JSR at P=500 bar, T=500-1100 K, Φ=0.5, and different dilution ratios (D=30% and 90%).

The impacts of real-fluid effects on NH$_3$ oxidation are clearly illustrated in Fig. 6 at both dilution ratios. The impacts are seemingly affected by the dilution ratio. As such, the dependence of real-fluid effects on dilution ratios is further quantified. Figure 7 presents the normalized change in simulated mole fractions after incorporating real-fluid effects at three different dilution ratios (i.e., 30%, 60%, and 90%). It is clear from Fig. 7 that for most species, the real-fluid effect is limited and less influenced by dilution ratios at relatively low and very high temperatures around T=1100 K, as the oxidation is barely started and nearly completed, respectively (cf. Fig. 6). It can be inferred that the real-fluid effects are not only determined by the operation conditions but also by the oxidation stage (i.e. the degree of NH$_3$ consumption), as is illustrated in Section 3.2. Furthermore, a lower dilution ratio exhibits a more intensive real-fluid effect on reactants, accelerating the consumption of NH$_3$ and O$_2$. At lower dilution ratios, the real-fluid effect on products (e.g., N$_2$ and H$_2$O) is less pronounced, while the "reversion phenomenon" for NO production is weakened. For NO$_2$, real-fluid effects, which promote NO$_2$ production (Fig. 6), are less distinct at lower dilution ratios at around T<900 K, whereas an opposite trend is observed at around T>900 K.

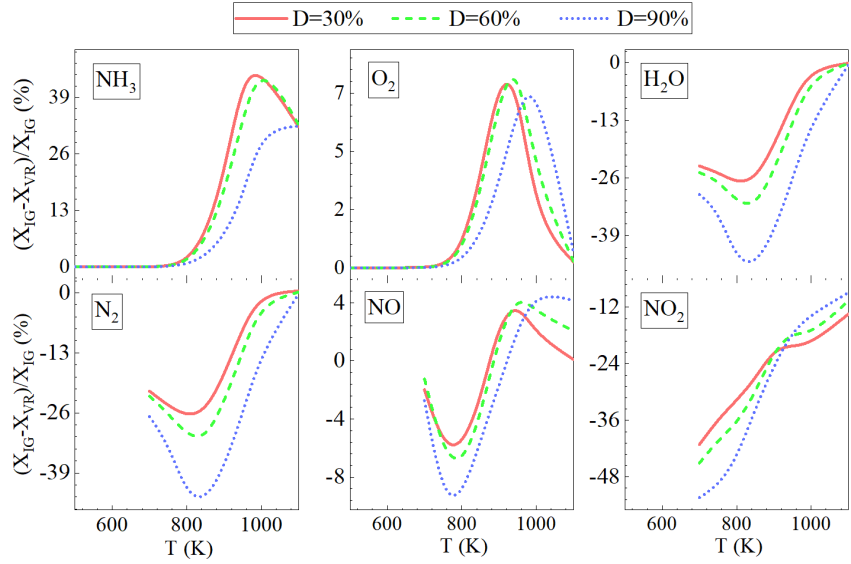

Figure 7. Deviation of simulated mole fractions during $NH_3$ oxidation (Case 3) based on high-order Virial EoS from those based on ideal EoS in a supercritical JSR at P=500 bar, T=500-1100 K, Φ=0.5, and different dilution ratios (D=30%, 60%, and 90%).

## 3.4 Influences of equivalence ratios

The influence of equivalence ratios on the real-fluid effect during $NH_3$ oxidation (i.e., Case 3 in Table 1) is further investigated, with normalized species profiles and normalized changes in calculated mole fractions shown in Fig. 8 and Fig. 9, respectively. Figure 8 illustrates that the fuel-lean condition distinctly advances and accelerates the oxidation process in comparison to the fuel-rich condition, thereby promoting the yield of $H_2O$ and $N_2$. At 1100 K, the oxidation is nearly completed under Φ=0.5 but halfway done at Φ=2. The real-fluid effects impose stronger influences at the fuel-lean condition, which is observed for all species.

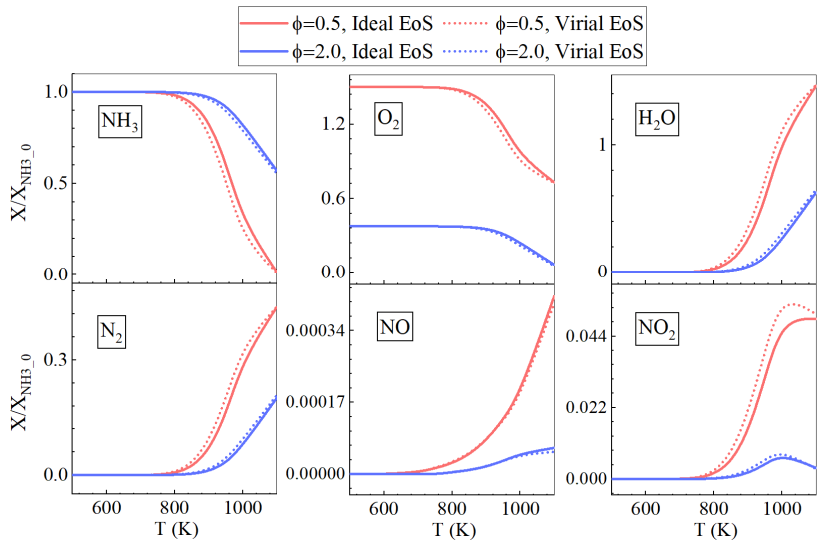

Figure 8. Normalized species profiles during $NH_3/O_2/CO_2$ oxidation calculated based on high-order Virial EoS and ideal EoS in a supercritical JSR at P=500 bar, T=500-1100 K, D=90% and different equivalence ratios (Φ=0.5 and 2).

The obvious change in real fluid effects, as observed in Fig. 8, warrants quantifications of real-fluid effects over wider equivalence ratios. This is also achieved by quantifying the relative change in species mole fractions after incorporating the real-fluid behaviors. The results are summarized in Fig. 9. As can be seen in Fig. 9, the increasing trends of normalized changes with higher temperatures are more monotonical for all species investigated under a rich condition. In contrast, the lean condition, i.e., Φ=0.5, witnesses a non-monotonic trend in the real-fluid effects from low to high temperatures. Compared to the fuel-rich condition, the fuel-lean condition considerably augments the real-fluid effects on $NH_3$ oxidation, while suppressing the real-fluid effects on the formation of NO and $NO_2$, especially at higher temperatures. A lean condition weakens the "reversion phenomenon" for NO. Overall, low equivalence ratios and low dilution ratios show similar trends of real-fluid effects.

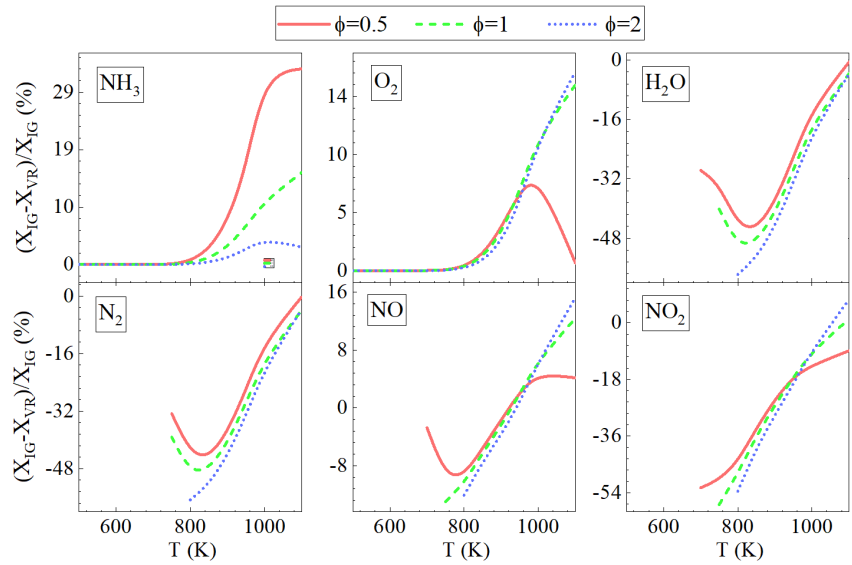

Figure 9. Deviation of simulated mole fractions during $NH_3$ oxidation (Case 3) based on high-order Virial EoS from those based on ideal EoS in a supercritical JSR at P=500 bar, T=500-1100 K, D=90% and different equivalence ratios (Φ=0.5, 1, and 2).

## 3.5 Real-fluid effects on fuel blending characteristics

As can be seen from Fig. 4, $NH_3$ is the most difficult to oxidize (oxidation starts at the highest temperature), while $H_2$ and $CH_4$ show relatively higher reactivities and are expected to promote the oxidation of $NH_3$ when blended with $NH_3$ [6]. In this regard, it is also necessary to investigate the impact of real-fluid behaviors on the blending characteristics between $NH_3$ and $H_2$ or $CH_4$. Figure 10 summarizes the normalized species profiles during the oxidation of pure $NH_3$ (BR=0%), a 50% $NH_3/H_2$ blend (BR=50%), and a $NH_3/CH_4$ blend (BR=50%) at 500 K<T<1100 K. The results show that adding $H_2$ and $CH_4$ significantly advances the onset of $NH_3$ oxidation. Notably, at all conditions investigated in this study, fuel blending with $H_2$ and $CH_4$ cannot reduce $NO_x$ emissions. A previous numerical study [44] observed a similar phenomenon at BR<80% for fuel blending with $NH_3$ and $H_2$. The diverse real-fluid effects on fuel blending characteristics are also obvious in Fig. 10. Specifically, the real-fluid effects advance the oxidation of $NH_3$ in the $H_2/NH_3$ mixture, while retarding the oxidation of $NH_3$ in the $CH_4/NH_3$ mixture. This opposite trend is also observed for $N_2$ and NO. It is the most obvious with NO species profiles that the real-fluid effects inhibit and promote NO production for the $CH_4/NH_3$ and $H_2/NH_3$ blends, respectively.

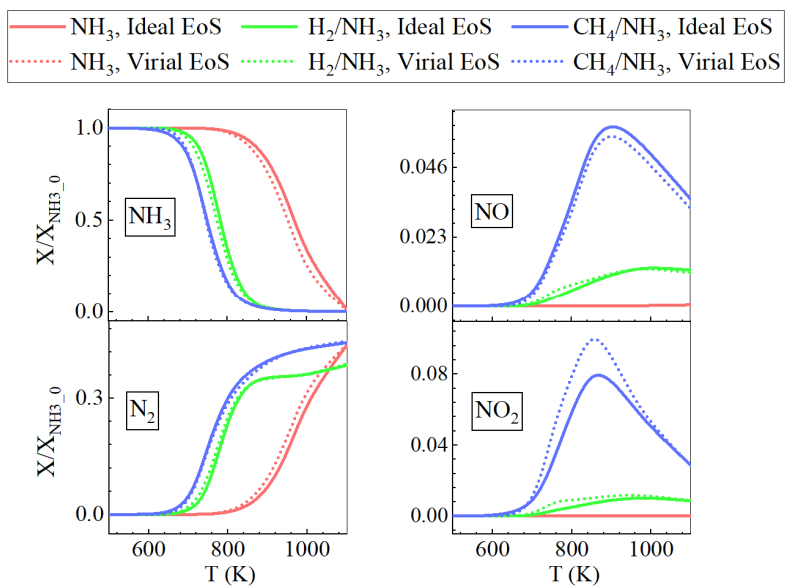

**Figure 10.** Normalized species profiles during the oxidation of $NH_3$ (BR=0%, Case 3) and 50% $NH_3/H_2$ or $NH_3/CH_4$ blends (BR=50%, Cases 4 and 5 respectively) calculated based on high-order Virial EoS and ideal EoS in a supercritical JSR at P=500 bar, T=500-1100 K, D=90% and Φ=0.5. BR – blending ratio.

To better illustrate the diverse change in real-fluid effects for different blends, the normalized change in simulated mole fractions with real-fluid behaviors incorporated is quantified over the studied temperature range for more species. The results are summarized in Fig. 11. As the addition of $H_2$ or $CH_4$ accelerates the reaction process, the oxidation is completed at round T>950 K (cf. Fig. 10), where the real-fluid effect for all species is limited and is weaker than that without fuel addition (cf. Fig. 11). However, at lower temperatures around 650 K<T<850 K, the oxidation is ongoing for fuel blend $NH_3/H_2$ while it is still at the initial stage for fuel $NH_3$. The real-fluid effect on $O_2$ reaches the lowest at BR=100%. As the real-fluid effect on the oxidation, which is at the initial stage, is limited (also observed in Figs. 5, 7, and 9), the real-fluid effect on all species during $NH_3/H_2$ oxidation (cf. Fig. 11(a)) is more distinct at 650 K<T<850 K than that of the fuel $NH_3$. By blending with $H_2$, the real-fluid effect increases the NO and $NO_2$ emissions by 85% and 165% at T=750 K, respectively. However, at BR=0%, these values are only 9% and 52%, respectively. Figure 11(b) witnesses a different real-fluid effect on the $CH_4$ blending behavior. The real-fluid effect on reactants $NH_3$ and $O_2$ is similar to that of fuels $NH_3$ and $NH_3/H_2$, but the real-fluid effect is less pronounced for all components at relatively low and moderate temperatures. The real-fluid effect on $H_2O$ at BR=50% is weaker than that at BR=0% at all temperatures. In addition, the real-fluid effect suppresses $N_2$ production and NO production at all temperatures.

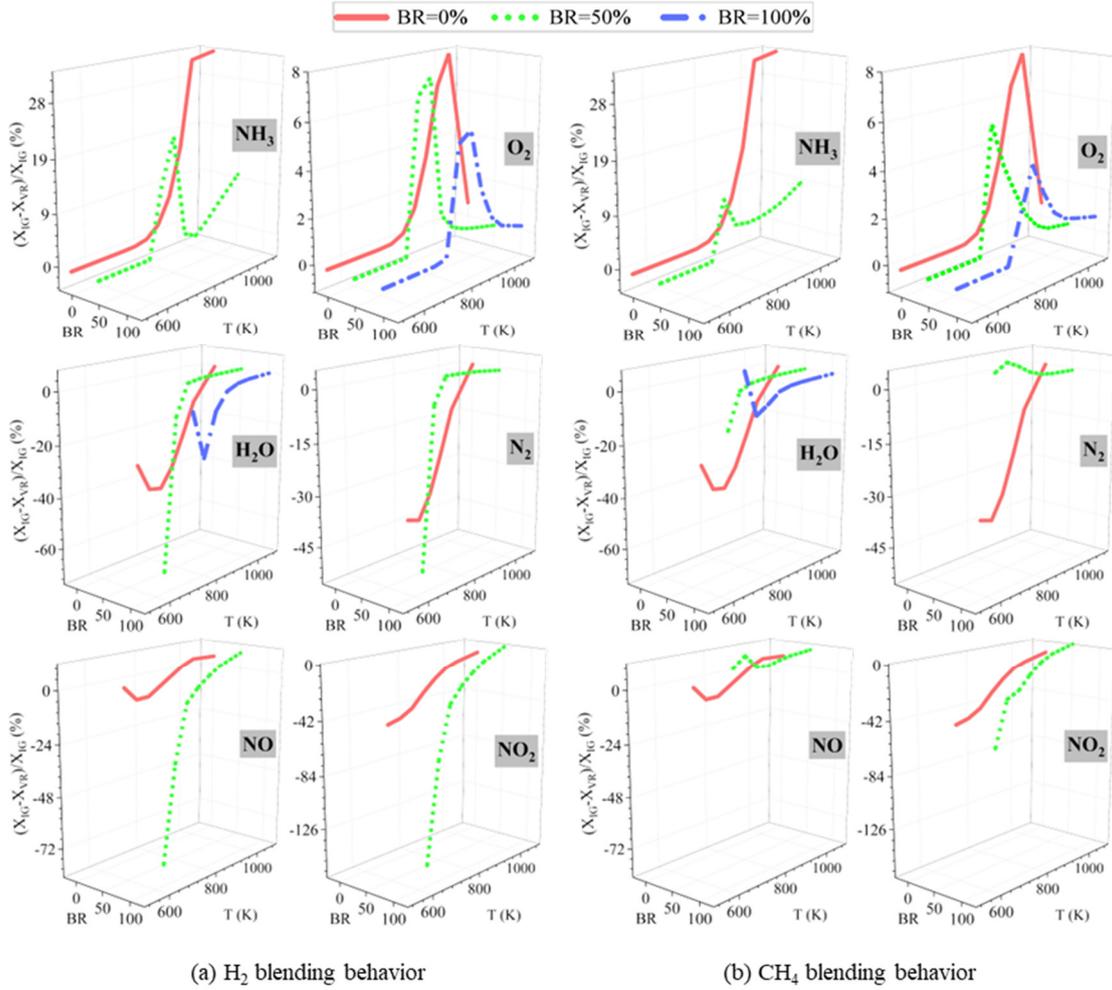

Figure 11. Deviation of simulated mole fractions during the oxidation of $NH_3$ (BR=0%, Case 3), 50% $NH_3/H_2$ or $NH_3/CH_4$ blends (BR=50%, Cases 4 and 5 respectively), and $H_2$ or $CH_4$ (BR=100%, Cases 1 and 2 respectively) based on high-order Virial EoS from those calculated based on ideal EoS in a supercritical JSR at P=500 bar, T=500-1100 K, D=90% and Φ=0.5. BR – blending ratio.

Similarly, fuel blending behaviors are investigated at different pressures P=100-1000 bar and constant temperature T=900 K, with normalized species profiles shown in Fig. 12. High pressures promote oxidation reactivity and accelerate the consumption of $NH_3$. The mole fractions of $NH_3$ and $N_2$ feature a more intensive dependence on pressure for fuel $NH_3$ than for the other two fuel blends. It is also clear from Fig. 12 that real-fluid effects impose opposite impacts on the oxidation characteristics for the formation of NO, where the production of NO is inhibited while promoted by the real-fluid effects for $CH_4/NH_3$ and $H_2/NH_3$ blends, respectively. Another interesting observation from Fig. 12 is the diminished influences from fuel blending after incorporating the real-fluid effects. For instance, with ideal gas assumption, the oxidation of $NH_3$ is accelerated with $CH_4$ or $H_2$ blending by a large change at the lowest pressure condition in Fig. 12. However, as pressure increases to 1000 bar and the real-fluid effects are enhanced considerably, this promoting effect from fuel blending is reduced by nearly 40%. The quantitative influences of real-fluid effects on fuel blending behaviors are further shown in Fig. S3, where the shifts in real-fluid effects across different fuel blends are clearly seen at 900 K. These shifts could imply that fuel blending strategies aiming to improve $NH_3$ combustion performance might become less effective at high-pressure conditions under the influences of real-fluid effects, which should be taken into account when developing advanced NH$_3$ combustion strategies via numerical approaches.

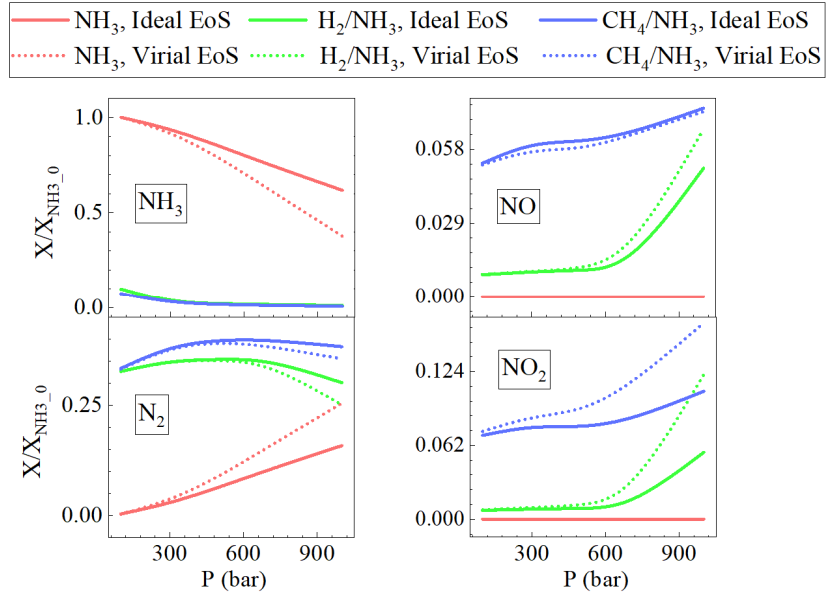

**Figure 12.** Normalized species profiles during the oxidation of NH$_3$ (BR=0%, Case 3) and 50% NH$_3$/H$_2$ or NH$_3$/CH$_4$ blends (BR=50%, Cases 4 and 5 respectively) calculated based on high-order Virial EoS and ideal EoS in a supercritical JSR at T=900 K, P=100-1000 bar, D=90% and Φ=0.5. BR – blending ratio.

The results in Figs. 4 – 12 also highlight the need to fully incorporate real-fluid effects when validating chemical kinetic mechanisms against high-pressure JSR experiments. Without this consideration, the error introduced in simulated species mole fractions can be as high as ±85%, as can be seen from Fig. 5. These errors are greater than the typical level of measurement uncertainties in JSR, and propagation of such levels of error to chemical kinetic mechanisms can disqualify the developed chemical kinetic mechanisms for any meaningful modeling work.

## 3.6 Sensitivity analysis

To gain an in-depth knowledge of the real-fluid effect on species profiles, sensitivity analysis is conducted for NH$_3$, NO, and NO$_2$ during NH$_3$/O$_2$/CO$_2$ oxidation (i.e., Case 3 in Table 1) at T=900 K and 1100 K. The reaction sensitivity coefficient $S_j$ with respect to mole fractions, $X$, is calculated by perturbating the reaction rate constant, $k_j$, by ±10%, as shown in Eq. 23. The results are summarized in Fig. 13, covering the top 13 most sensitive reactions. A positive (negative) value indicates that a reaction suppresses (promotes) consumption or promotes (suppresses) production for the species investigated.

$$S_j = \frac{ln(X^+/X^-)}{ln(k_j^+/k_j^-)} \tag{23}$$

For NH$_3$ in Fig. 13(a), sensitivity coefficients for all reactions are smaller at 900 K than at 1100 K. For both results using ideal EoS and Virial EoS, reaction pathways of NH$_2$ with NO and NO$_2$ are the most dominant reactions. For both pathways, a chain propagating/branching product channel,

$$NH_2 + NO = NNH + OH \qquad (R1)$$
$$NH_2 + NO_2 = H_2NO + NO \qquad (R2)$$

competes with a respective chain terminating channel,

$$NH_2 + NO = N_2 + H_2O \qquad (R3)$$
$$NH_2 + NO_2 = N_2O + H_2O \qquad (R4)$$

with the R1 and R2 promoting $NH_3$ consumption while R3 and R4 inhibit $NH_3$ consumption. Another inhibiting reaction $H_2NO+OH=HNO+H_2O$ (R5) is also impactful at 1100 K but exhibits no sensitivity at 900 K. Additionally, the propagating reaction $NH_2 + HO_2 = H_2NO + OH$ (R6) has nearly the same sensitivity coefficient at 900 K and 1100 K. The real-fluid effect tends to augment sensitivity for reactions R1-6, with a more distinct enhancement at 1100 K than at 900 K. This explains the stronger promotion of the real-fluid effect on $NH_3$ consumption at higher temperatures than at lower temperatures, as observed in Figs. 5, 7, and 9. In addition, the real-fluid effects exhibit an overall higher enhancement on the promoting reactions than the inhibiting reactions. As such, the real-fluid effects increase system reactivity and accelerate the oxidation process for $NH_3$.

For NO, as shown in Fig. 13(b), R1, R2, and R6 promote its production, while R3 and R4 suppress its production. At T=900 K, although the real-fluid effects strengthen the contribution from R4, the inhibiting effect from R3, which is the most inhibiting reaction on NO production, is weakened by the real-fluid effects. These changes are expected to lead to an overall enhanced NO production, which agrees with trends observed in Fig. 5. However, at 1100 K, the contributions from R1 and R6 become negligible, while those from R3 and R4 preserve. The promoting effects from R1, R2, and R6 at this temperature weaken and even reverse the influences of real-fluid effects on NO production, which is also confirmed in Fig. 5, where real-fluid effects shift from promoting to inhibiting NO production. The trends for $NO_2$ (Fig. 13(c)) at T=900 K are similar to those for NO at the same temperature. As such, the real-fluid effects also promote $NO_2$ production. However, at 1100 K, the contributions from R2 and R6 become negligible, leading to a reduced promoting effect of real-fluid effects on $NO_2$ production.

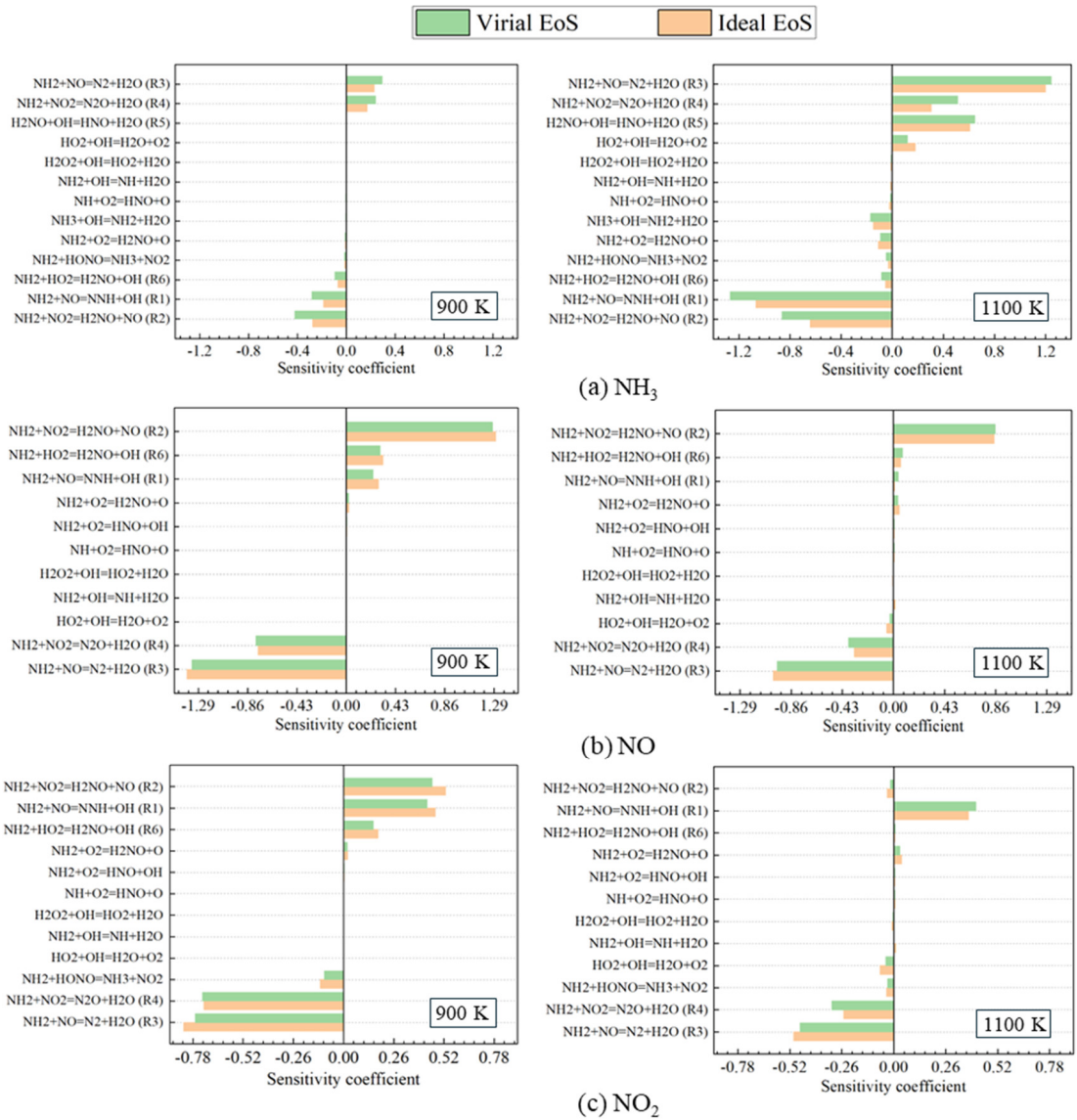

**Figure 13.** Sensitivity analysis on the mole fractions of (a) $NH_3$, (b) NO, and (c) $NO_2$ during oxidation of mixture $NH_3/O_2/CO_2$ (Case 3) based on ideal EoS and high-order Virial EoS in a supercritical JSR at P=500 bar, D=90%, Φ=0.5 and two different temperatures (i.e., 900 K and 1100 K).

# 4. Conclusions

This study presents, for the first time, a novel framework coupling high-order Virial EoS, *ab initio* intermolecular potentials, and real-fluid governing equations to characterize the real-fluid effects during supercritical oxidation in JSRs. This has not been achievable in the past as all previous modeling efforts in JSR have assumed ideal gas behavior, although this assumption might fail at high pressure conditions. Through comprehensive simulations, the real-fluid effects on $NH_3$ oxidation and $NH_3$ blending effects with $H_2$ and $CH_4$ are quantified at wide ranges of temperature, pressure, equivalence ratio and dilution ratio. Detailed analyses of the results indicate the following:

(1) The real-fluid effects promote $NH_3$ oxidation by advancing the onset of oxidation toward lower temperatures, with enhanced effects at higher temperatures, higher pressures, lower dilution ratios, and lower equivalence ratios. The production and consumption of NO and $NO_2$ are considerably influenced by real-fluid effects, with reversed effects observed when temperature increases.

(2) Fuel blending effects with $H_2$ and $CH_4$ are significantly complicated by the real-fluid effects, where the real-fluid effects tend to diminish the positive influences from fuel blending.

(3) Sensitivity analysis with and without considering real-fluid effects highlights the important and diverse influences of real-fluid effects on the most sensitive pathways for $NH_3$ oxidation, primarily via shifting the competition in the $NH_2 + NO$ and $NH_2 + NO_2$ pathways.

(4) Without considering real-fluid effects, the error introduced in simulated species mole fractions can be as high as $\pm 85\%$ at the conditions studied, highlighting the necessity of adequately incorporating real-fluid effects in future studies that aim to develop or validate chemical kinetic mechanisms against high-pressure JSR experiments.

# Acknowledgments

The work described in this paper is supported by the Research Grants Council of the Hong Kong Special Administrative Region, China under PolyU P0046985 for ECS project funded in 2023/24 Exercise, the Otto Poon Charitable Foundation under P0050998, the National Natural Science Foundation of China under 52406158, the Chief Executive's Policy Unit of HKSAR under the Public Policy Research Funding Scheme (2024.A6.252.24B), and the Natural Science Foundation of Guangdong Province under 2023A1515010976 and 2024A1515011486.

# Author Contributions

M.W. and S.C. conceived and designed the framework and performed the calculations, with technical inputs from R.T., X.R., H.W., T.Z., and D.Y.. All authors contributed to writing the paper.

# Declaration of Competing Interests

The authors declare no competing interests.